\documentclass[prd,amsmath,amssymb,longbibliography,superscriptaddress,twocolumn,nofootinbib,10pt]{revtex4-1}

\pdfoutput=1
\usepackage{graphicx}
\usepackage{dcolumn}
\usepackage{bm}
\usepackage{amssymb}
\usepackage{latexsym}
\usepackage{booktabs}
\usepackage{amsmath}
\usepackage{multirow}
\usepackage{array}

\usepackage[colorlinks=true, linkcolor=blue, citecolor=blue]{hyperref}

\begin{document}

\title{Measuring cosmic curvature with non-CMB observations}

\author{Peng-Ju Wu}\email{wupengju@nxu.edu.cn}
\affiliation{School of Physics, Ningxia University, Yinchuan 750021, China}

\author{Xin Zhang}\thanks{Corresponding author}
\email{zhangxin@mail.neu.edu.cn}
\affiliation{Department of Physics, College of Sciences, Northeastern University, Shenyang 110819, China}
\affiliation{Key Laboratory of Data Analytics and Optimization for Smart Industry (Ministry of Education), Northeastern University, Shenyang 110819, China}
\affiliation{National Frontiers Science Center for Industrial Intelligence and Systems Optimization, Northeastern University, Shenyang 110819, China}

\begin{abstract}
The cosmic curvature $\Omega_{K}$ is an important parameter related to the inflationary cosmology and the ultimate fate of the universe. In this work, we adopt the non-CMB observations to constrain $\Omega_{K}$ in the $\Lambda$CDM model and its extensions. The DESI baryon acoustic oscillation, DES type Ia supernova, cosmic chronometer, and strong gravitational lensing time delay data are considered. We find that the data combination favors an open universe in $\Lambda$CDM, specifically $\Omega_{K}=0.106\pm0.056$ at the $1\sigma$ level, which is in $2.6\sigma$ tension with the Planck CMB result supporting our universe being closed. In the $\Lambda$CDM extensions, the data combination is consistent with a flat universe. It is noteworthy that when the cosmic chronometer data is excluded, the derived constraints demonstrate increased statistical preference for an open universe. Specifically, $\Omega_{K}=0.146\pm0.060$ in $\Lambda$CDM, which is in $3.1\sigma$ tension with the CMB result, and the flat universe scenario can be ruled out at $>1\sigma$ level in the $\Lambda$CDM extensions. The constraining power mainly stems from DESI's distance measurements. Given the advantages of the full-shape power spectrum in $\Omega_{K}$ measurements and the support for a closed universe from previous full-shape analyses, it is necessary to validate our findings with DESI's full-shape data. We adopt the Akaike information criterion to compare different cosmological models. The result shows that non-flat models fit the observational data better than the flat $\Lambda$CDM model, which indicates that flat $\Lambda$CDM may not be the ultimate model of cosmology.
\end{abstract}

\maketitle
\section{Introduction}\label{sec1}
Whether our universe is spatially open, flat or closed is a fundamental issue in cosmology. A non-zero curvature would have profound implications for the primordial inflation paradigm and the ultimate fate of the universe. Measuring the sign and value of $\Omega_{K}$ is of great significance for understanding the evolution of the universe and the nature of dark energy. The question has gained a lot of interest in the past few years, particularly in light of the Planck cosmic microwave background (CMB) observations \cite{Aghanim:2018eyx,Park:2017xbl,Handley:2019tkm,DiValentino:2019qzk,Efstathiou:2020wem}. It is found that the CMB data alone favor a closed universe, $\Omega_{K}=-0.044_{-0.015}^{+0.018}$ \cite{Aghanim:2018eyx}. Spatial flatness is an indicator of inflation \cite{Anselmi:2022uvj}. If the universe is not flat, this would cast serious doubt on the possibility that inflation could have happened. This deviation from the flat universe is interpreted as the undetected systematics or new physics beyond the standard model of cosmology, i.e., the $\Lambda$ cold dark matter ($\Lambda$CDM) model.

The ability of CMB data to constrain $\Omega_{K}$ is limited by the geometrical degeneracy which can be broken by including other observational data \cite{Zaldarriaga:1997ch,Efstathiou:1998xx,Bond:1997wr,Vagnozzi:2020rcz,Vagnozzi:2020dfn,Dhawan:2021mel}. For instance, the CMB data combined with the baryon acoustic oscillation (BAO) measurements give $\Omega_{K}=0.0007\pm0.0019$, suggesting that our universe is flat to within the $1\sigma$ confidence level \cite{Aghanim:2018eyx}. This is inconsistent with the result of the CMB data alone. Some people expressed doubts about the soundness of such combinations, believing that there may be a ``curvature tension'' in the current data \cite{Handley:2019tkm}. Of course, there are different voices. Efstathiou and Gratton claimed that the CMB data are consistent with a flat universe \cite{Efstathiou:2020wem}. Whether this tension exists is still an open question, and recent developments even offer hints for an open universe \cite{DESI:2024mwx,Jiang:2024xnu}. It should be pointed out that in addition to the possible curvature tension, there are other measurement inconsistencies that do exist between early- and late-universe observations, such as the measurements for the Hubble constant $H_0$ and the amplitude of the matter power spectrum (see Ref.~\cite{Verde:2019ivm} for a review). Remarkably, the $H_0$ value measured by the Cepheid-supernova distance ladder is in above $5\sigma$ tension with that inferred from the CMB observation assuming $\Lambda$CDM \cite{Riess:2021jrx}. When combining observations for parameter constraints, it is necessary to consider the measurement inconsistencies between the data. The existence of these tensions reduces the rationality of combining CMB with late-universe probes for measuring the curvature parameter.

The motivation of the present work is to constrain the curvature parameter using only the late-universe observations. The mainstream late-time probes include BAO (standard ruler), type Ia supernova (SN, standard candle) and cosmic chronometer (CC, standard clock). For a long time before, these non-CMB observations could not well constrain the curvature parameter. However, the situation may have changed. Recently, the Dark Energy Spectroscopic Instrument (DESI) collaboration released the high-precision BAO data based on the precise observations of galaxies, quasars, and Lyman-$\alpha$ forests \cite{DESI:2024uvr,DESI:2024lzq}. Earlier, the Dark Energy Survey (DES) program published the high-quality samples of SN Ia discovered during its five-year operation \cite{DES:2024tys}. Furthermore, more than 30 CC data have been available for cosmological parameter inference so far \cite{Moresco:2022phi}. We will utilize the data from these three probes. To further tighten the constraints, we will also consider the strong gravitational lensing time delay (TD) observations \cite{Suyu:2009by,Jee:2019hah,Suyu:2013kha,Chen:2019ejq,Wong:2016dpo,Birrer:2018vtm,Rusu:2019xrq,DES:2019fny,Agnello:2017mwu}. These four probes observe the universe from different perspectives, and combining them is expected to break cosmological parameter degeneracies, thereby narrowing the constraints.

Currently, many curvature measurements are achieved after making assumptions about the nature of dark energy, i.e., dark energy behaves like a cosmological constant $\Lambda$ with an equation of state (EoS) of $w=-1$. However, the observational data leaves room for dark energy EoS to deviate from $-1$, which weakens the persuasiveness of the $\Omega_{K}$ constraints assuming $\Lambda$CDM, since $\Omega_{K}$ and $w$ are strongly degenerate. In addition, many people believe that $\Lambda$CDM is not the ultimate model of cosmology. On one hand, it has some theoretical problems \cite{Weinberg:1988cp,Sahni:1999gb}; on the other hand, the CMB results for $\Lambda$CDM are in tension with some late-universe observations~\cite{Verde:2019ivm}. Cosmologists have conceived many theories beyond it to solve the problems and reconcile the tensions \cite{Guo:2018ans}. It is necessary to measure the curvature parameter in those extended models. In this work, we shall consider some typical extensions to the $\Lambda$CDM model, mainly for the evolutionary behavior of dark energy. We shall also consider the possible interaction between dark energy and dark matter. This paper focuses on measuring the cosmic curvature in $\Lambda$CDM and its extensions using the non-CMB observations, and study the impact of various extensions on the $\Omega_{K}$ constraints.

The remainder of this paper is organized as follows. We briefly describe the methodology in Sec.~\ref{sec2}. Sec.~\ref{sec3} contains the observational data we adopted. We present the results and make some discussions in Sec.~\ref{sec4}. Finally, we give our conclusions in Sec.~\ref{sec5}.

\section{methodology}\label{sec2}
If space is homogeneous and isotropic, the spacetime can be described by the Friedmann-Lema\^\i tre-Robertson-Walker (FLRW) metric with line element
\begin{align}\label{FLRW}
{\rm d} s^{2}=-c^2{\rm d} t^{2}+a^{2}(t)\left[\frac{{\rm d} r^{2}}{1-K r^{2}}+r^{2}({\rm d}\theta^{2}+\sin ^{2} \theta {\rm d} \phi^{2})\right],
\end{align}
where $c$ is the speed of light, $a$ is the scale factor, and $K$ is the spatial curvature, which is related to the curvature parameter by $\Omega_{K} = -Kc^2 / H_0^2$, with $H_0$ being the Hubble constant. Then $\Omega_{K}>0$, $\Omega_{K}=0$ and $\Omega_{K}<0$ correspond to the spatially open, flat and closed universe, respectively.

Distance measure is the most direct way to understand the evolution of the universe. Next, we introduce the definitions of various cosmological distances. Throughout this paper, we adopt $D_{\rm C}(z)$, $D_{\rm M}(z)$, $D_{\rm L}(z)$, $D_{\rm A}(z)$, and $D_{\rm H}(z)$ to represent the comoving distance, transverse comoving distance, luminosity distance, angular diameter distance, and Hubble distance, respectively. The comoving distance is defined as
\begin{equation}\label{DC}
D_{\rm C}(z)=\int_{0}^{z} \frac{c{\rm d}z'}{H(z')}=\frac{c}{H_0}\int_{0}^{z} \frac{{\rm d}z'}{E(z')},
\end{equation}
where $H(z)$ is the Hubble parameter and $E(z)=H(z)/H_0$ is the dimensionless Hubble parameter. The transverse comoving distance is given by \cite{Hogg:1999ad}
\begin{equation}\label{DM}
D_{\rm M}(z)=\left\{
\begin{aligned}
&\frac{c}{H_0\sqrt{\Omega_K}}\sinh{\left[\frac{H_0\sqrt{\Omega_K}}{c}D_{\rm C}(z)\right]} & {\rm if}\ \Omega_{K}>0 ,  \\
&D_{\rm C}(z)                                                                             & {\rm if}\ \Omega_{K}=0 , \\
&\frac{c}{H_0\sqrt{-\Omega_K}}\sin{\left[\frac{H_0\sqrt{-\Omega_K}}{c}D_{\rm C}(z)\right]}  & {\rm if}\ \Omega_{K}<0 .
\end{aligned}
\right.
\end{equation}
The luminosity distance, angular diameter distance, and Hubble distance are defined as
\begin{align}\label{DLDADH}
D_{\rm L}=D_{\rm M}\cdot(1+z);\ D_{\rm A}=D_{\rm M}/(1+z);\  D_{\rm H}=c/H.
\end{align}

By measuring distances, we can constrain cosmological models. For the $\Lambda$CDM model, the dimensionless Hubble parameter can be written as
\begin{align}\label{LCDMEz}
E(z)=\sqrt{\Omega_{\rm m}(1+z)^3+\Omega_K(1+z)^2+\Omega_{\Lambda}},
\end{align}
where $\Omega_{\rm m}$ and $\Omega_{\Lambda}$ refer to the density parameters of non-relativistic matter and dark energy, respectively. In this model, the dark energy EoS $w$ (the ratio of pressure to density) is $-1$. Now we turn to extensions to the $\Lambda$CDM cosmology. We consider four dynamical dark energy models, which are (\romannumeral1) the $w$CDM model with a constant EoS $w(z)=w$; (\romannumeral2) the exponential model with an evolving EoS $w(z)=w\exp[z/(1 + z)]/(1 + z)$ \cite{Yang:2018qmz}; (\romannumeral3) the CPL model with an evolving EoS $w(z)=w_0+w_a z/(1+z)$ \cite{Chevallier:2000qy,Linder:2002et}; (\romannumeral4) the JBP model with an EoS of the form $w(z)=w_0+w_a z/(1+z)^2$, which was proposed to solve the high--$z$ issues within the CPL model \cite{Jassal:2004ej}. The first two models are one-parameter extensions ($w$), and the last two models are two-parameter extensions ($w_0$ and $w_a$) to the $\Lambda$CDM model. Certainly, there are other reasonable parameterizations for dark energy, such as the MZ parameterization model \cite{Ma:2011nc}, among others. However, in this paper, it is not possible for us to go through all of these models; we can only select a few as typical representatives.

We also consider a coupling between dark energy and dark matter. If there exists an interaction between them, the energy conservation equations can be written as
\begin{align}\label{IDEQ}
\dot{\rho}_{\rm de}&=-3H(1+w){\rho}_{\rm de}+Q, \nonumber \\
\dot{\rho}_{\rm c}&=-3H{\rho}_{\rm c}-Q,
\end{align}
where ${\rho}_{\rm de}$ and ${\rho}_{\rm c}$ represent the energy densities of dark energy and cold dark matter, respectively, the dot denotes the derivative with respect to the cosmic time $t$, $w$ is the dark energy EoS for which we take $-1$, and $Q$ is the energy transfer rate. In this paper, we employ a phenomenological form of $Q=\beta H \rho_{\rm de}$, where $\beta$ is a dimensionless coupling parameter; $\beta>0$ indicates that dark matter decays into dark energy, $\beta<0$ denotes that dark energy decays into dark matter, and $\beta=0$ means that there is no interaction between them. Of course, for the form of interaction, we can also consider other possible forms, but the latest results indicate that the most recent observations, including those from DESI, support this form of interaction to about $3\sigma$ level \cite{Li:2024qso}. Therefore, in this paper, we take this form as a typical representative to consider interacting dark energy (IDE).
Here, the IDE model is treated as a one-parameter extension ($\beta$) to $\Lambda$CDM. For more detailed introduction to IDE models, see Refs.~\cite{Valiviita:2008iv,Koyama:2009gd,Clemson:2011an,Li:2013bya,Li:2014eha,Li:2014cee,Li:2015vla,Xia:2016vnp,Zhang:2017ize,Guo:2017deu,Gao:2021xnk,Gao:2022ahg,Li:2023fdk}.

We will explore the spatial geometry of the universe in the $\Lambda$CDM model and its five extensions using the latest observational data.

\section{Observational data}\label{sec3}
In this work, we adopt the Markov Chain Monte Carlo method to maximize the likelihood $L\propto\rm{exp}(-\chi^2/2)$ to infer the probability distributions of cosmological parameters using the observational data. The $\chi^2$ function of each dataset can be written as $\Delta\boldsymbol{D}^T\boldsymbol{C}^{-1}\Delta\boldsymbol{D}$, where $\Delta\boldsymbol{D}$ is the vector of observable residuals representing the difference between observation and theory, and $\boldsymbol{C}$ is the covariance matrix. Next, we introduce the data utilized in this work.

$\bullet$ {\bf Type Ia Supernovae.} The SN Ia can be used as a standard candle to measure the luminosity distance. We consider the DES sample of 1829 distinct SNe~Ia covering $0.025<z<1.3$ \cite{DES:2024tys}. This sample quintuples the number of high-quality SNe beyond $z > 0.5$ compared to the previous leading compilation of Pantheon+. For an SN~Ia, the distance modulus is given by $\mu=m_{B}-M_{B}$, where $m_{B}$ is the observed magnitude in the rest-frame $B$-band and $M_{B}$ is the absolute magnitude. The theoretical distance modulus is defined as
\begin{align}
\mu_{\rm th}(z)=5 \log \left[\frac{D_{\rm L}(z)}{\mathrm{Mpc}}\right]+25.
\end{align}
The observed distance moduli and errors, as well as the covariance between different data points can be found at the website.\footnote{\url{https://github.com/des-science/DES-SN5YR}}

$\bullet$ {\bf Baryon Acoustic Oscillations.}
The BAO scale provides us a standard ruler to measure the angular diameter distance and Hubble parameter. We consider the recently released DESI BAO data and summarize them in Table~\ref{BAO}. The data points with same redshifts are correlated (for example the $D_{\rm M}/r_{\rm d}$ and $D_{\rm H}/r_{\rm d}$ measurements are correlated at $z=0.51$), and the covariance matrices can be found in corresponding papers \cite{DESI:2024uvr,DESI:2024lzq,DESI:2024mwx} and website.\footnote{\url{https://data.desi.lbl.gov/doc/releases/}} When constructing the $\chi^2$ function for MCMC analysis, we will take into account the correlations between different observables at each point. The total $\chi^2$ function equals the sum of the $\chi^2$ function at each data point.
\begin{table}
\renewcommand\arraystretch{1}
\caption{The BAO measurements from DESI. Here, $r_{\rm d}$ is the sound horizon and $D_{\rm V}(z)=\left[D_{\rm M}(z)\right]^{2/3}\left[cD_{\rm H}(z)\right]^{1/3}$ is the comoving volume-averaged distance.}
\label{BAO}
\centering
\setlength{\tabcolsep}{5pt}
\begin{tabular}{l c c c}
\bottomrule[1pt]
& Redshift $z$  &Observable                     & Value                     \\
\bottomrule[1pt]
& 0.30          & $D_{\rm V}/r_{\rm d}$         & $7.93\pm0.15$             \\
& 0.51          & $D_{\rm M}/r_{\rm d}$         & $13.62\pm0.25$            \\
& 0.51          & $D_{\rm H}/r_{\rm d}$         & $20.98\pm0.61$            \\
& 0.71          & $D_{\rm M}/r_{\rm d}$         & $0.497\pm0.045$           \\
& 0.71          & $D_{\rm H}/r_{\rm d}$         & $13.38\pm0.18$            \\
& 0.93          & $D_{\rm M}/r_{\rm d}$         & $22.43\pm0.48$            \\
& 0.93          & $D_{\rm H}/r_{\rm d}$         & $0.459\pm0.038$           \\
& 1.32          & $D_{\rm M}/r_{\rm d}$         & $17.65\pm0.30$            \\
& 1.32          & $D_{\rm H}/r_{\rm d}$         & $19.78\pm0.46$            \\
& 1.49          & $D_{\rm V}/r_{\rm d}$         & $0.473\pm0.041$           \\
& 2.33          & $D_{\rm M}/r_{\rm d}$         & $19.5\pm1.0$              \\
& 2.33          & $D_{\rm H}/r_{\rm d}$         & $19.6\pm2.1$              \\
\bottomrule[1pt]
\end{tabular}
\end{table}
In the present work, we treat the sound horizon scale $r_{\rm d}$ as a free parameter in cosmological constraints, thereby ensuring that our results remain independent of both the early-universe observational priors and the computational modeling of the sound horizon at the drag epoch (i.e., model-independent). In the cosmological constraints reported by the DESI collaboration, the product of the sound horizon and dimensionless Hubble constant $r_{\rm d}h$ was adopted as a composite sampling parameter when using the BAO data alone. In this paper, we implement separate sampling for both $H_0$ and $r_{\rm d}$ as individual free parameters. We have compared the constraint results derived from these two sampling methods and found that the $\Omega_K$ constraints are consistent. It should be pointed out that some studies have underscored the substantial benefits of full-shape power spectrum analyses in constraining the cosmic curvature (see, e.g., Refs.~\cite{Glanville:2022xes,Chudaykin:2020ghx}). Nevertheless, as DESI has not yet released full-shape data, we defer the exploration of such analysis to future works.

\begin{table}
\renewcommand\arraystretch{1}
\caption{The 32 $H(z)$ measurements obtained with the CC method.}
\label{CCHz}
\centering
\setlength{\tabcolsep}{5pt}
\begin{tabular}{l c c c}
\bottomrule[1pt]
& Redshift $z$ & $H(z)\ [\,\rm km/s/Mpc\,]$            & Reference                       \\
\bottomrule[1pt]
& 0.07         & 69  $\pm$      19.6                  & \citep{Zhang:2012mp}             \\
& 0.09         & 69    $\pm$   12                    & \citep{Simon:2004tf}             \\
& 0.12         & 68.6  $\pm$26.2                  & \citep{Zhang:2012mp}             \\
& 0.17         & 83   $\pm$ 8                     & \citep{Simon:2004tf}             \\
& 0.179        & 75  $\pm$ 4                     & \citep{Moresco:2012jh}           \\
& 0.199        & 75   $\pm$ 5                     & \citep{Moresco:2012jh}           \\
& 0.2          & 72.9  $\pm$ 29.6                  & \citep{Zhang:2012mp}             \\
& 0.27         & 77     $\pm$14                    & \citep{Simon:2004tf}             \\
& 0.28         & 88.8  $\pm$36.6                  & \citep{Zhang:2012mp}             \\
& 0.352        & 83    $\pm$ 14                    & \citep{Moresco:2012jh}           \\
& 0.38         & 83    $\pm$ 13.5                  & \citep{Moresco:2016mzx}          \\
& 0.4          & 95     $\pm$ 17                    & \citep{Simon:2004tf}             \\
& 0.4004       & 77    $\pm$ 10.2                  & \citep{Moresco:2016mzx}          \\
& 0.425        & 87.1  $\pm$ 11.2                  & \citep{Moresco:2016mzx}          \\
& 0.445        & 92.8  $\pm$ 12.9                  & \citep{Moresco:2016mzx}          \\
& 0.47         & 89   $\pm$ 49.6                  & \citep{Ratsimbazafy:2017vga}     \\
& 0.4783       & 80.9   $\pm$ 9                     & \citep{Moresco:2016mzx}          \\
& 0.48         & 97    $\pm$ 62                    & \citep{Stern:2009ep}             \\
& 0.593        & 104   $\pm$ 13                    & \citep{Moresco:2012jh}           \\
& 0.68         & 92    $\pm$ 8                     & \citep{Moresco:2012jh}           \\
& 0.75         & 98.8  $\pm$ 33.6                  & \citep{Borghi:2021rft}           \\
& 0.781        & 105   $\pm$ 12                    & \citep{Moresco:2012jh}           \\
& 0.875        & 125   $\pm$ 17                    & \citep{Moresco:2012jh}           \\
& 0.88         & 90    $\pm$ 40                    & \citep{Stern:2009ep}             \\
& 0.9          & 117    $\pm$ 23                    & \citep{Simon:2004tf}             \\
& 1.037        & 154   $\pm$ 20                    & \citep{Moresco:2012jh}           \\
& 1.3          & 168    $\pm$ 17                    & \citep{Simon:2004tf}             \\
& 1.363        & 160   $\pm$33.6                  & \citep{Moresco:2015cya}          \\
& 1.43         & 177    $\pm$ 18                    & \citep{Simon:2004tf}             \\
& 1.53         & 140   $\pm$ 14                    & \citep{Simon:2004tf}             \\
& 1.75         & 202    $\pm$ 40                    & \citep{Simon:2004tf}             \\
& 1.965        & 186.5  $\pm$ 50.4                  & \citep{Moresco:2015cya}          \\
\bottomrule[1pt]
\end{tabular}
\end{table}
$\bullet$ {\bf Cosmic Chronometers.} The CC method provides a direct way to measure the Hubble parameter. In a universe described by the FLRW metric, the Hubble parameter can be written in terms of the differential time evolution of the universe $\Delta t$ in a given redshift interval,
\begin{align}
\label{CC}
H(z)=-\frac{1}{1+z} \frac{\Delta z}{\Delta t}.
\end{align}
The best CCs are extremely massive and passively evolving galaxies. By measuring these galaxies, one can obtain their redshifts and differences in age, and thus achieving the estimation of $H(z)$. We summarize the latest 32 CC $H(z)$ measurements in Table~\ref{CCHz}. Among them, 15 data points from Refs.~\cite{Moresco:2012jh,Moresco:2016mzx,Moresco:2015cya} are correlated, and the covariance matrix can be found in the website\footnote{\url{https://gitlab.com/mmoresco/CCcovariance}}. Note that the CC data are indeed affected by systematic uncertainties that have not yet been fully controlled \cite{Kjerrgren:2021zuo} (such as the age-dating systematics and modeling of stellar populations), and thus they are not widely adopted by the cosmological community. For this reason, we shall present constraint results separately for ``with CC'' and ``without CC'' cases.

$\bullet$ {\bf Strong Gravitational Lensing.} Strong gravitational lensing (SGL) is a rare astronomical phenomenon. For an SGL system, the source, lens and observer are well aligned, so multiple images of the source appear to the observer. Moreover, the photons from source corresponding to different images travel through different spacetime paths, which makes delays between the arrival times. The TD between images depends on the gravitational potential and so-called time-delay distance which is calculated by
\begin{align}
\label{SGL-TD}
D_{\Delta t}\equiv(1+z_{\rm l})\displaystyle{\frac{D_{\rm l}D_{\rm s}}{D_{\rm ls}}},
\end{align}
where $D_{\rm l}$, $D_{\rm s}$, and $D_{\rm ls}$ are the angular diameter distances between observer and lens, between observer and source, and between lens and source, respectively. We summarize the existing SGL systems with $D_{\Delta t}$ and $D_{\rm l}$ measurements in Table~\ref{TD}. For more details, we refer readers to the website.\footnote{\url{https://zenodo.org/records/3633035}} For simplicity, the error of each data point will be treated in a Gaussian form in the present work.

\begin{table}
\renewcommand\arraystretch{1.2}
\caption{The time-delay distances and angular diameter distances for seven TD measurements. Here, $z_{\rm l}$ and $z_{\rm s}$ are the redshifts of lens and source respectively. Note that the error of each data is treated in a Gaussian form in this work.}
\label{TD}
\setlength{\tabcolsep}{5pt}
\begin{tabular}{l  c c c c c}
\bottomrule[1pt]
& $z_{\rm l}$   & $z_{\rm s}$   & $D_{\Delta t}\ [\rm Mpc]$     & $D_{\rm l}\ [\rm Mpc]$    & References                                      \\
\bottomrule[1pt]
& 0.6304        & 1.394         & $5156^{+296}_{-236}$          & $1228^{+177}_{-151}$      & \cite{Suyu:2009by,Jee:2019hah}                 \\
& 0.295         & 0.654         & $2096^{+98}_{-83}$            & $804^{+141}_{-112}$       & \cite{Suyu:2013kha,Chen:2019ejq}               \\
& 0.4546        & 1.693         & $2707^{+183}_{-168}$          & $-$                       & \cite{Wong:2016dpo,Chen:2019ejq}               \\
& 0.745         & 1.789         & $5769^{+589}_{-471}$          & $1805^{+555}_{-398}$      & \cite{Birrer:2018vtm}                          \\
& 0.6575        & 1.662         & $4784^{+399}_{-248}$          & $-$                       & \cite{Rusu:2019xrq}                            \\
& 0.311         & 1.722         & $1470^{+137}_{-127}$          & $697^{+186}_{-144}$       & \cite{Chen:2019ejq}                            \\
& 0.597         & 2.375         & $3382^{+146}_{-115}$          & $1711^{+376}_{-280}$      & \cite{DES:2019fny,Agnello:2017mwu}             \\
\bottomrule[1pt]
\end{tabular}
\end{table}

\section{Results and discussions}\label{sec4}
In this section, we report the constraint results from the combination of late-universe probes. We perform the MCMC analysis to infer the probability distributions of cosmological parameters. The parameters we sample include $H_0$, $\Omega_{\rm m}$, $\Omega_K$, $r_{\rm d}$, $w$, $w_0$, $w_a$, and $\beta$. We take flat priors for them with ranges of $H_0\in[20,\,100]\,\rm km/s/Mpc$, $\Omega_{\rm m}\in[0,\,1]$, $\Omega_K\in[-0.3,\,0.3]$, $r_{\rm d}\in[100,\,200]\,\rm{Mpc}$, $w\in[-3,\,1]$, $w_0\in[-3,\,1]$, $w_a\in[-5,\,5]$, and $\beta\in[-2,\,2]$. We measure the convergence of the MCMC chains by checking that all parameters have $R-1<0.01$, where $R$ is the potential scale reduction factor of the Gelman-Rubin diagnostics.

\begin{figure}
\includegraphics[scale=0.7]{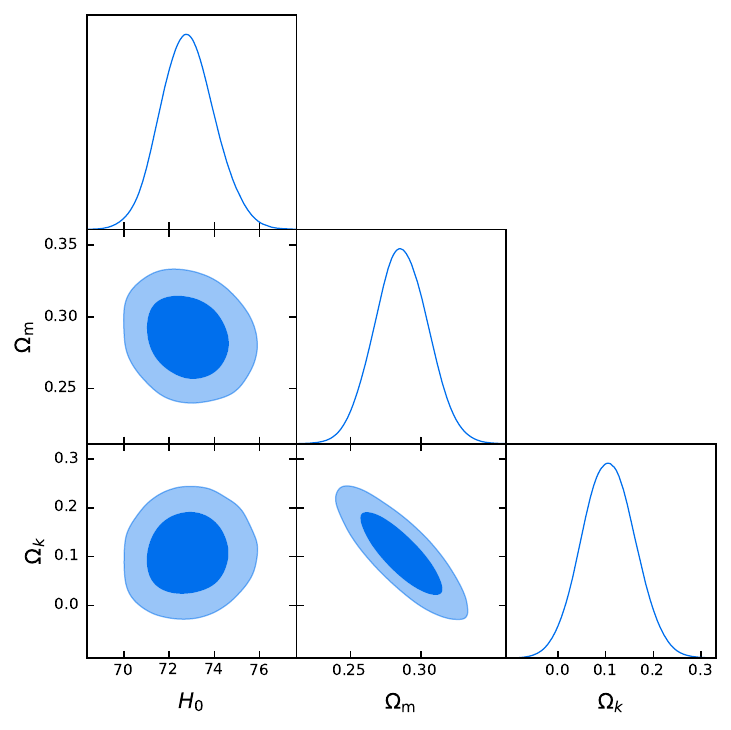}
\centering
\caption{Constraints on the non-flat $\Lambda$CDM model using the BAO+SN+CC+TD data.}
\label{LCDM}
\end{figure}
\begin{table*}[!htb]
\caption{Constraints on the cosmological parameters in the $\Lambda$CDM, $w$CDM, Exp, CPL, JBP and IDE models using the data combination of BAO+SN+CC+TD. The cosmic curvature is treated as a free parameter in all these models. The factor $\Delta$AIC refers to the Akaike information criterion difference between extended models and the flat $\Lambda$CDM model.}
\label{tab:results}
\setlength{\tabcolsep}{2mm}
\renewcommand{\arraystretch}{1.4}
\begin{center}{\centerline{
\begin{tabular}{cccccccc}
\bottomrule[1pt]
Model   & $H_0\,[\,\rm km/s/Mpc\,]$     &$\Omega_{\rm m}$       &$\Omega_{K}$                   &$w$ or $w_0$                           &$w_a$                                  &$\beta$        &$\Delta{\rm AIC}$  \\
\bottomrule[1pt]
{$\Lambda$CDM}
        &$72.8^{+1.1}_{-1.3}$           &$0.286\pm0.019$                &$0.106\pm0.056$                &$-$                                        &$-$                                        &$-$                    &$-1.786$            \\
{$w$CDM}
        &$71.9\pm1.3$                   &$0.290\pm0.019$                &$0.023\pm0.073$                &$-0.887^{+0.065}_{-0.051}$      &$-$                                       &$-$                    &$-3.086$       \\
{Exp}
        &$71.9\pm1.3$                   &$0.263\pm0.019$                &$0.018\pm0.073$                &$-0.846^{+0.068}_{-0.056}$     &$-$                                        &$-$                    &$-2.800$       \\
{CPL}
        &$71.3\pm1.3$                   &$0.317\pm0.023$                &$0.054\pm0.075$                &$-0.721^{+0.093}_{-0.11}$          &$-1.57^{+0.88}_{-0.67}$        &$-$                    &$-5.552$          \\
{JBP}
        &$71.1\pm1.4$                   &$0.312\pm0.022$                &$0.050\pm0.073$               &$-0.64^{+0.12}_{-0.14}$                &$-2.5^{+1.4}_{-1.1}$           &$-$                        &$-6.031$      \\
{IDE}
        &$72.0\pm1.3$                   &$0.387\pm0.029$                &$0.039\pm0.075$                &$-$                                        &$-$                                        &$-0.39\pm0.14$     &$-3.374$          \\
\bottomrule[1pt]
\end{tabular}}}
\end{center}
\end{table*}

We initiate the investigation by employing the combination of four datasets. For comparison, we first constrain the flat $\Lambda$CDM model and the results are $H_0=72.7\pm1.2\,\rm km/s/Mpc$ and $\Omega_{\rm m}=0.317\pm0.011$. The matter density parameter $\Omega_{\rm m}$ is consistent with that inferred from the Planck CMB data, but the Hubble constant value is in $4.1\sigma$ tension with the CMB result. If the curvature parameter is taken as a free parameter in MCMC analysis, the constraint contours shown in Figure~\ref{LCDM} can be obtained. As can be seen, the data combination strongly supports an open universe, specially $\Omega_{K}=0.106\pm0.056$ at the $1\sigma$ confidence level (C.L.). According to Ref.~\cite{Aghanim:2018eyx}, the CMB data alone favor a closed universe, i.e., $\Omega_{K}=-0.044_{-0.015}^{+0.018}$. Our result is in around $2.6\sigma$ tension with the CMB result. Importantly, the signs of the two $\Omega_{K}$ values are different, that is, they stand in complete opposition. Note that some people claimed that the CMB data are consistent with a flat universe \cite{Efstathiou:2020wem}, but even that is inconsistent with our result here. When considering $\Omega_K$ in $\Lambda$CDM constraints, the combination of late-time observations offers $H_0=72.8^{+1.1}_{-1.3}\,\rm km/s/Mpc$ and $\Omega_{\rm m}=0.286\pm0.019$, while the CMB data provide $H_0=57.4\pm3.3\,\rm km/s/Mpc$ and $\Omega_{\rm m}=0.434^{+0.044}_{-0.054}$. We can see that the early- and late-time observations are inconsistent in measuring all three important parameters $H_0$, $\Omega_{\rm m}$ and $\Omega_K$. All of these imply that $\Lambda$CDM may not be the ultimate model of cosmology and there may be new physics beyond it.

\begin{figure}
\includegraphics[scale=0.5]{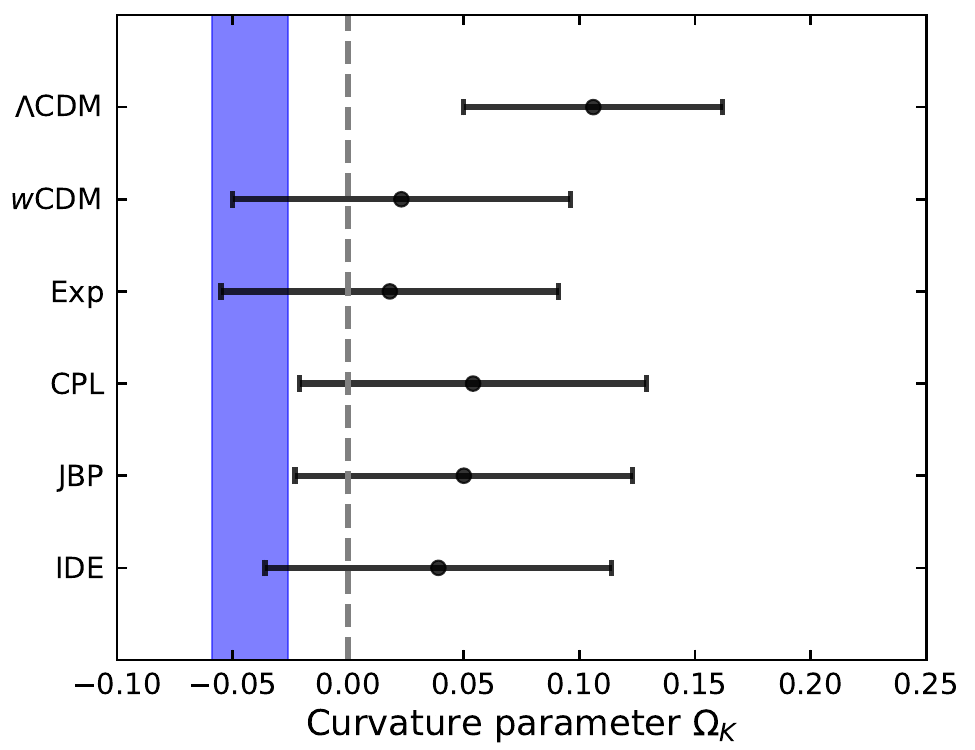}
\centering
\caption{Constraints on the curvature parameter in $\Lambda$CDM and its extensions using the BAO+SN+CC+TD data. The blue shaded region denotes the CMB result assuming $\Lambda$CDM and the grey vertical dashed line corresponds to the spatially flat universe.}
\label{extensions}
\end{figure}

The above analysis is for the $\Lambda$CDM model. In recent decades, various extensions to $\Lambda$CDM have been attempted to solve its inherent problems or reconcile the current measurement inconsistencies. It is important to measure the curvature parameter in these extended models and examine what insights the late-universe probes can provide. We constrain $\Omega_K$ in some $\Lambda$CDM extensions mentioned above. The $1\sigma$ errors for the marginalized parameter constraints are summarized in Table~\ref{tab:results}. For clarity, we plot the constraints on $\Omega_K$ in Figure~\ref{extensions}. The grey vertical dashed line corresponds to the spatially flat universe, and the blue band represents the $1\sigma$ C.L. interval obtained from the Planck CMB observation assuming $\Lambda$CDM. As can be seen, the constraint results are consistent with a spatially flat unverse for all $\Lambda$CDM extensions. However, regardless of whether the dark energy EoS evolves or not, and whether there is an interaction between dark energy and dark matter, the central value of $\Omega_{K}$ is positive and has a significant deviation from zero (ranging from $0.018$ to $0.054$), that is, the data combination still has some support for an open universe, although not up to the $1\sigma$ C.L. To some extent, our conclusions are robust, as all models tend to favor an open universe over a closed one.

Now we turn to the comparison of the models based on their fittings to the observational data. For $\Lambda$CDM extensions, a model with more parameters tends to produce a smaller $\chi^2$. Therefore, the $\chi^2$ comparison is unfair for comparing models. In this work, we adopt the Akaike information criterion (AIC) to compare different models. The AIC is defined as $\chi^2_{\rm min}+2k$, where $k$ refers to the number of free parameters. A model with a smaller AIC is more favored by observations. We adopt the relative value of the AIC between different models, $\Delta{\rm AIC}=\Delta\chi^2_{\rm min}+2\Delta k$, to complete our analysis. In this work, the flat $\Lambda$CDM model serves as a reference model. Generally, compared to the reference model, a model with $\Delta{\rm AIC}<2$ is substantially supported, a model with $4<\Delta{\rm AIC}<7$ is less supported, and a model with $\Delta{\rm AIC}>10$ is essentially not supported. We calculate the $\Delta{\rm AIC}$ values and present them in the last column of Table~\ref{tab:results}.

As can be seen, when the BAO+SN+CC+TD data are employed, the $\Delta{\rm AIC}$ value for the non-flat $\Lambda$CDM model is negative, which indicates a better fitting to the observations than the flat $\Lambda$CDM model. When the dark energy EoS is not $-1$, but a constant $w$ that can be some other value, $\Delta{\rm AIC}$ becomes more negative. The central value of EoS parameter for both the $w$CDM and Exp models lies within the quintessence-like regime ($w>-1$), and the cosmological constant falls out the $2\sigma$ C.L. regions. When the dark energy EoS is dynamically evolving, $\Delta$AIC becomes further negative, indicating that they can better fit the observational data. In addition, the model with an interaction between dark energy and dark matter (dark energy transforms into dark matter here), also shows superiority than the $\Lambda$CDM model. All these models perform better than flat $\Lambda$CDM, indicating that their increased model complexities are statistically supported when fitting the BAO+SN+CC+TD data. Overall, the dynamically evolving dark energy models are most supported by observations. The results add strong evidence to the argument that there are new physics beyond $\Lambda$CDM and the standard model of cosmology should be extended.

\begin{figure}
\includegraphics[scale=0.65]{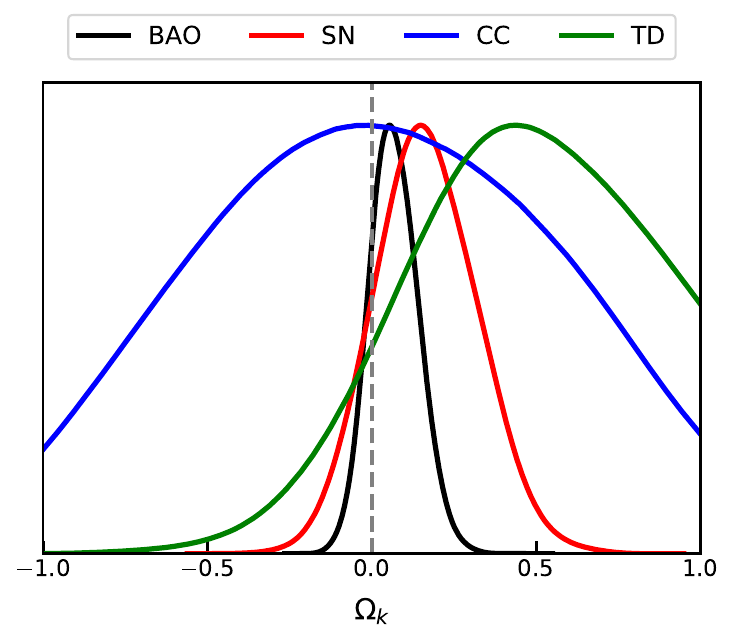}
\centering
\caption{The constraints on $\Omega_K$ in the $\Lambda$CDM model using the latest BAO, SN, CC and TD data, respectively.}
\label{LCDM-each}
\end{figure}

\begin{table*}[!htb]
\caption{Constraints on the cosmological parameters in the $\Lambda$CDM, $w$CDM, Exp, CPL, JBP and IDE models using the data combination of BAO+SN+TD. The cosmic curvature is treated as a free parameter in all these models. The factor $\Delta$AIC refers to the Akaike information criterion difference between extended models and the flat $\Lambda$CDM model.}
\label{tab:results1}
\setlength{\tabcolsep}{2mm}
\renewcommand{\arraystretch}{1.4}
\begin{center}{\centerline{
\begin{tabular}{cccccccc}
\bottomrule[1pt]
Model   & $H_0\,[\,\rm km/s/Mpc\,]$ &$\Omega_{\rm m}$   &$\Omega_{K}$       &$w$ or $w_0$                       &$w_a$                          &$\beta$                    &$\Delta{\rm AIC}$  \\
\bottomrule[1pt]
{$\Lambda$CDM}
        &$74.8\pm1.5$               &$0.282\pm0.020$        &$0.146\pm0.060$        &$-$                                        &$-$                                &$-$                        &$-4.377$            \\
{$w$CDM}
        &$74.0\pm1.7$               &$0.284\pm0.019$        &$0.093\pm0.079$        &$-0.931^{+0.084}_{-0.058}$     &$-$                                &$-$                        &$-3.511$       \\
{Exp}
        &$74.0\pm1.7$               &$0.262\pm0.019$        &$0.091\pm0.081$        &$-0.895^{+0.089}_{-0.064}$     &$-$                                &$-$                        &$-3.198$       \\
{CPL}
        &$73.3\pm1.8$               &$0.307\pm0.023$        &$0.118\pm0.080$        &$-0.75\pm0.12$                     &$-1.70^{+1.0}_{-0.80}$    &$-$                         &$-5.342$          \\
{JBP}
        &$73.1\pm1.8$               &$0.306\pm0.023$        &$0.109\pm0.081$        &$-0.66\pm0.16$                     &$-2.7^{+1.6}_{-1.2}$       &$-$                        &$-5.387$      \\
{IDE}
        &$74.1\pm1.7$               &$0.402\pm0.030$        &$0.106\pm0.082$        &$-$                                        &$-$                                &$-0.41\pm0.15$     &$-4.131$          \\
\bottomrule[1pt]
\end{tabular}}}
\end{center}
\end{table*}
The above results are based on the synergy of four probes. Whether each probe supports the open universe is also worth exploring. Taking the $\Lambda$CDM model as an example, we adopt the four probes to constrain the curvature parameter respectively, and the results are shown in Figure~\ref{LCDM-each}. We can see that the central value of $\Omega_K$ derived from the BAO, SN and TD is obviously greater than zero. The center value obtained from the CC data is close to zero. Concretely, the marginalized constraints for $\Omega_K$ are $0.068^{+0.068}_{-0.080}$, $0.16\pm0.16$, $0.43^{+0.43}_{-0.27}$ and $0.01\pm0.48$ for the BAO, SN, TD and CC data, respectively. The error obtained from the CC data is the largest, due to the relatively poor quality of the data. In general, there are no obvious inconsistencies in the $\Omega_K$ measurements from these four late-universe probes. As can be seen, the constraining power primarily stems from the DESI BAO data. This work directly utilizes DESI's distance measurements. However, several studies have indicated that employing full-shape power spectrum offers significant advantages for measuring the cosmic curvature. Previously released full-shape data from other galaxy surveys tend to support a closed universe \cite{Glanville:2022xes,Chudaykin:2020ghx}. Consequently, we plan to employ DESI's upcoming full-shape data to conduct a rigorous examination of the conclusions drawn in this work.

\begin{figure}
\includegraphics[scale=0.5]{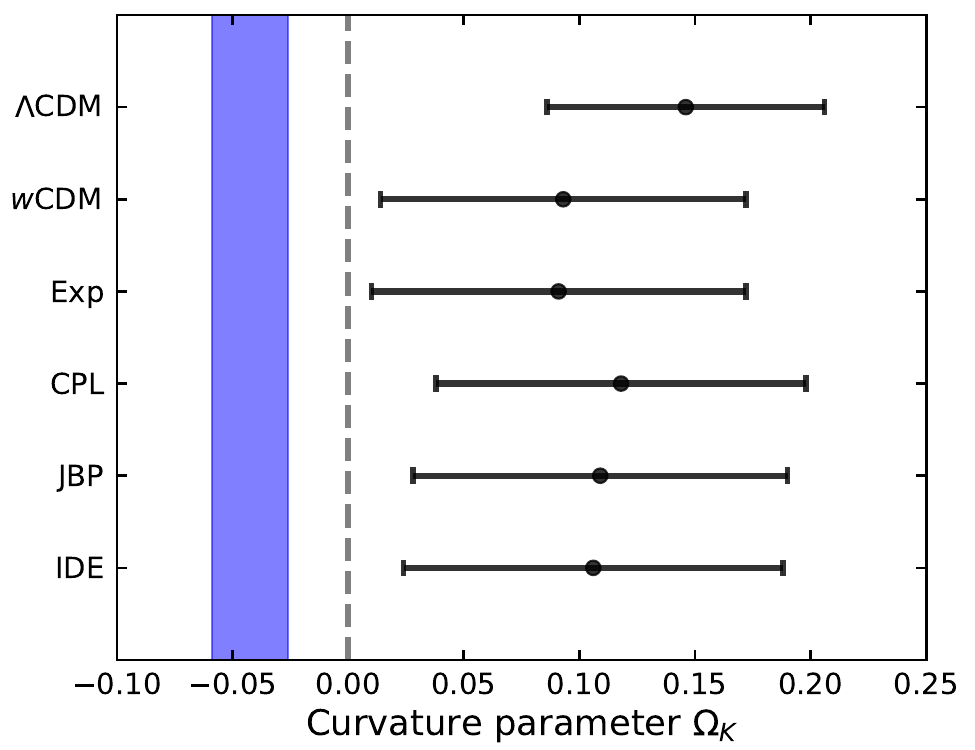}
\centering
\caption{Constraints on the curvature parameter in $\Lambda$CDM and its extensions using the BAO+SN+TD data. The blue shaded region denotes the CMB result assuming $\Lambda$CDM and the grey vertical dashed line corresponds to the flat universe.}
\label{extensions1}
\end{figure}

It should be stressed that the CC data may not constitute a reliable source of information considering some concerns \cite{Kjerrgren:2021zuo}, and thus they are not widely adopted by the cosmological community. The CC data may bias the derived constraints on key parameters, such as $H_0$, $\Omega_{\rm m}$, $\Omega_{K}$ and $w$. We re-evaluate the cosmic curvature by excluding the CC data, with the updated results summarized in Table~\ref{tab:results1}. A marked discrepancy emerges between the new constraints and previous findings. Most notably, both the central values of the Hubble constant and the curvature parameter exhibit a statistically significant increase under the revised analysis framework. The updated constraints demonstrate a stronger preference for an open universe, and we plot the $\Omega_K$ constraints from the BAO+SN+TD data  in Figure~\ref{extensions1}. The BAO+SN+TD data provide  $\Omega_{K}=0.146\pm0.060$ at the $1\sigma$ C.L. in the $\Lambda$CDM model, which is in around $3.1\sigma$ tension with the Planck CMB result. In addition, the flat universe scenario can be excluded at $>1\sigma$ C.L. in all $\Lambda$CDM extensions. The AIC analysis shows that all non-flat models perform better than the flat $\Lambda$CDM model, indicating that their increased model complexities are statistically supported when fitting the BAO+SN+TD data.

Currently, in the $\Lambda$CDM cosmology, the combination of four late-time probes clearly supports an open universe. However, due to the degeneracy between the dark energy EoS and curvature parameter as well as the limited constraint power of the observational data, we cannot put tight constraints on $\Omega_K$ in $\Lambda$CDM extensions. The large constraint error of $\Omega_K$ leaves the window open for a flat or even closed universe. Nonetheless, the central value of $\Omega_K$ is positive and has a significant deviation from zero for all extended models. Considering the currently uncontrolled systematic uncertainties in the CC data, we also measure the cosmic curvature by excluding the CC data. We find that although the obtained confidence interval is enlarged, the central value of $\Omega_{K}$ exhibits a more significant deviation from zero. The derived constraints more strongly favor an open universe, and the flat universe scenario can be excluded at $>1\sigma$ C.L. in all extended models of $\Lambda$CDM. To determine the spatial curvature of the universe at high confidence levels in $\Lambda$CDM extensions, we need more high-quality observational data. In addition to the traditional cosmological probes utilized in this paper, we also look forward to the performance of emerging probes in cosmological parameter estimation \cite{Wu:2022dgy,Wu:2022jkf,Kumar:2024bvp}. We will return to this issue when the amount of observational data increases.

\section{Conclusions}\label{sec5}
The cosmic curvature $\Omega_{K}$ relates to the inflationary paradigm and the ultimate fate of the universe. Knowing whether the universe is open, flat, or closed is crucial for us to understand its evolution and the nature of dark energy. In this work, we adopt four late-time cosmological probes to measure $\Omega_{K}$ in the $\Lambda$CDM model and its five extensions. The latest baryon acoustic oscillation, type Ia supernova, cosmic chronometer and strong gravitational lensing time delay data are employed.

In the $\Lambda$CDM model, the synergy of four probes supports an open universe, specifically $\Omega_{K}=0.106\pm0.056$ at the $1\sigma$ C.L., which is in around $2.6\sigma$ tension with the CMB result favoring that our universe is spatially closed. We find that there are no inconsistencies between the measurements of $\Omega_K$ from each late-universe probe. In all $\Lambda$CDM extensions, the data combination of four probes is consistent with a flat universe. Due to the degeneracy between the dark energy equation of state and curvature parameter as well as the limited constraint ability of the data, we cannot put tight constraints on $\Omega_{K}$. However, the central value of $\Omega_{K}$ is positive and deviates far from zero, indicating that the data still have some support for an open universe, although the confidence level does not reach $1\sigma$. At the very least, we can assert that the data combination favor an open universe over a closed one.

Considering that the CC data may not constitute a reliable source of information due to some concerns, we also constrain the cosmic curvature by excluding the CC data. We find that the BAO+SN+TD data demonstrate a stronger preference for an open universe. Specially, they provide  $\Omega_{K}=0.146\pm0.060$ at the $1\sigma$ C.L. in the $\Lambda$CDM model, which is in around $3.1\sigma$ tension with the CMB result. Although the obtained confidence interval is enlarged, the central value of $\Omega_{K}$ exhibits a more significant deviation from zero. For this reason, the flat universe scenario is excluded at $>1\sigma$ C.L. in all the $\Lambda$CDM extensions.

However, it must be emphasized that our constraining power on the cosmic curvature primarily originates from the BAO data. In this work, we utilized the DESI BAO data rather than the full-shape matter power spectrum. Studies have indicated that the full-shape data offers greater advantages for measuring the cosmic curvature, which therefore represents a limitation of the present study. Historical full-shape data from other galaxy surveys tend to favor a closed universe. This trend may reflect unresolved systematics or methodological differences rather than true evidence against the flat $\Lambda$CDM model. We reserve the option to employ the DESI's upcoming full-shape data to revisit the conclusions drawn here.

Currently, neither early- nor late-time observations support a spatially flat universe in the $\Lambda$CDM cosmology, and compelling evidence for cosmic non-flatness has been identified in the $\Lambda$CDM extensions when using the BAO+SN+TD data. In light of these results, we recommend considering the cosmic curvature as a free parameter in cosmological parameter constraints. In addition, we find that all non-flat models provide a better fitting to the observational data than the flat $\Lambda$CDM model, which means that flat $\Lambda$CDM may not be the ultimate model of cosmology.

\begin{acknowledgments}
We thank Tian-Nuo Li and Guo-Hong Du for fruitful discussions. This work was supported by the National SKA Program of China (Grants Nos. 2022SKA0110200 and 2022SKA0110203), the National Natural Science Foundation of China (Grants Nos. 12473001, 11975072, and 11835009), and the National 111 Project (Grant No. B16009).

\end{acknowledgments}

\bibliography{curvature}

\end{document}